\documentclass[useAMS,usenatbib]{mn2e}
\usepackage{pstricks,wasysym,stmaryrd,enumerate,longtable,lscape,epsfig}
\usepackage{longtable,fancyhdr}
\usepackage{floatflt,graphicx}
\usepackage[T1]{fontenc}
\usepackage{times}
\newcommand{\dg}{$^{\circ}$}
\newcommand{\DOT}{.}

\newcommand\kms{km~s$^{-1}$}

\title[Sgr~A* at low radio frequencies: GMRT observations]{Sgr~A* at
low radio frequencies: GMRT observations}
\author[Subhashis Roy \& A. Pramesh Rao]
{Subhashis Roy\thanks{E-mail: roy@ncra.tifr.res.in} \& A. Pramesh
Rao\thanks{E-mail: pramesh@ncra.tifr.res.in} \\ National Centre for Radio
Astrophysics (TIFR), \\ Pune University Campus, Post Bag No.3, Ganeshkhind,
Pune 411 007, India}
\begin{document}
\date{}
\pagerange{L\pageref{firstpage}--L\pageref{lastpage}}
\pubyear{}

\maketitle

\label{firstpage}
\begin{abstract}

The central region of the Galaxy has been observed at 580, 620 and 1010 MHz
with the Giant Metrewave Radio Telescope (GMRT). We detect emission from
Sgr~A*, the compact object at the dynamical centre of the Galaxy, and
estimate its flux density at 620 MHz to be 0.5$\pm$0.1 Jy. This is the
first detection of Sgr~A* below 1 GHz \citep{IAU199.ROY,GC2003.ROY}, which
along with a possible detection at 330 MHz \citep{NORD2004} provides its
spectrum below 1 GHz.  Comparison of the 620 MHz map with maps made at other
frequencies indicates that most parts of the Sgr~A West HII region have
optical depth~$\sim$2.  However, Sgr~A*, which is seen in the same region in
projection, shows a slightly inverted spectral index between 1010 MHz and 620
MHz.  This is consistent with its high frequency spectral index, and indicates
that Sgr~A* is located in front of the Sgr~A West complex, and rules out any
low frequency turnover around 1 GHz, as suggested by \citet{DAVIES1976}.
\end{abstract}


\section{Introduction}
Being located two orders of magnitude closer than the nearest large
galaxy, the Galactic Centre (GC) can be studied at a much higher spatial
resolution and sensitivity than is possible for other galaxies. Because of
this advantage, we can identify unique objects like the Radio-arc
consisting of linear parallel filaments \citep{YUSEF-ZADEH1984}, or the
2.6$ \times 10^6$ M$_{\odot}$ black hole suggested to be associated with
the compact radio source Sgr~A* \citep{GHEZ1998}.
From the high resolution ($\sim$~arc~seconds) observation by the Very
Large Array (VLA) at radio frequencies \citep{EKERS1983,PEDLAR1989}, the
following sources within the central 15$'$ of the Galaxy have been
identified. \\
(i) At the dynamical centre of the Galaxy is the compact nonthermal radio
source known as Sgr-A*. (ii) Around Sgr~A* is the HII region Sgr-A West
\citep{EKERS1983}, whose morphology resembles a face-on spiral galaxy.
(iii) Near Sgr-A West is Sgr-A East, which is believed to be a supernova
remnant (SNR). (iv) A 7$'$ halo, which has been proposed to be a mixture
of thermal and non-thermal emission \citep{PEDLAR1989}.

Sgr~A* (see Melia \& Falcke 2001 for a review) was not detected at
frequencies below 960 MHz and observations at 408 MHz \citep{DAVIES1976}
and at 330 MHz \citep{PEDLAR1989} provide upper limits ($\le$0.1 Jy at 330
MHz) on its flux density. Sgr~A* probably has a low frequency turnover
below 1 GHz, but the nature of the turnover has never been examined in
detail \citep{MELIA2001}.  Recently, \citet{NORD2004} claim to have
detected Sgr~A* at 330 MHz.  However, the average brightness of the 7$'$ halo
seen towards Sgr~A* at 330 MHz is $\sim 100$ mJy/Beam (with the beamsize used
in their map), which is comparable to the claimed peak intensity of Sgr~A*
(Fig.~2 in Nord et al. 2004). The 7$'$ halo could be located in front of the
Sgr~A complex \citep{PEDLAR1989}, and presence of any small scale structure in
the halo along Sgr~A* can mimic its claimed detection. Therefore, detection of
Sgr~A* at 330 MHz remains provisional.

High resolution radio observations \citep{ROBERTS1993} show that Sgr~A~West
comprises of three major features of ionised gas known as Northern and Southern
arm and Western arc, which are embedded in a halo of lower density ionised gas
with an extent of about 1.5$'$ \citep{MEZGER1986, PEDLAR1989}.
Along the Northern arm, the gas appears to flow away from us. If this is
taken as an indication of gas falling in towards Sgr~A*, then this would
imply that the Northern arm is located in front of Sgr~A*.
Though \citet{WHITEOAK1983} have suggested Sgr~A* to be located in front
of Sgr A West, \citet{LISZT1983} detected HI absorption against Sgr~A* at
40$-$60 \kms, and not against Sgr~A~West. While \citet{LISZT1983}
attributes this discrepancy to the patchiness of the HI screen, the
relative location of Sgr~A* with respect to Sgr~A~West need to be
established.

Due to free-free absorption at low radio frequencies, HII regions tend to
get optically thick and absorption against another continuum object can be
used to constrain their relative location. 

We have observed the central half a degree region of the Galaxy at 580,
620 and 1010 MHz using the GMRT \citep{SWARUP1991}.
As a cross-check, we have observed Sgr~A* also at 580 MHz. To estimate its
spectral index between 1 GHz and 620 MHz and compare with its spectrum
obtained at higher radio frequencies, we have further observed this region
at 1010 MHz with the GMRT. In this paper, we will mainly discuss Sgr~A*
and Sgr~A~West HII region.
In Sect.~2, we describe the observations and data analysis.  The results
are presented in Sect.~3, and the inferences are discussed in Sect.~4.
The conclusions are presented in Sect.~5.

\section{Observations and data reduction}

The observations were carried out with the Giant Metrewave Radio Telescope
(GMRT) with a nominal bandwidth of 16 MHz on Aug 31 \& Sept 21, 2001 at 620 MHz
and in June 2002 at 580 MHz. Full synthesis ($\sim$7 hours) was performed at
these frequencies.  The field centre was set at RA (J2000)=17h46m07s, DEC
(J2000)=$-$28\dg57$'$02$^{''}$. The observations in the 1010 MHz band were
carried out on 9th of August 2003.
These were (1010 MHz) two snapshot observations each of about 45 minutes in
duration,
and the field centre was set at RA (J2000)=17h45m40.5s, DEC
(J2000)=$-$28\dg56$'$09$^{''}$. All these observations were carried out in the
spectral line mode with 128 frequency channels across 16 MHz. Absolute flux
density calibration  was performed using 3C48 and 3C286 data following
\citet{BAARS1977} scale.  The source 1751$-$253 and 1714$-$25 were used as
secondary and bandpass calibrator respectively. At 1010 MHz, 3C286 was used for
bandpass calibration.

The increase of $T_{sys}$ from the calibrator field to the target source
affects the source visibility amplitudes in the default observing mode
(i.e., the Automatic Level Control (ALC) in the system is turned on), and
we employed the following method to correct for the $T_{sys}$ variation.
With the ALC off, we estimated the ratio of the total power on the target
source and the calibrator 3C48 at 620 MHz. Since this ratio was quite
similar (within 10\%) for almost all the antennas, rather than multiplying
the antenna based gain, we multiplied the final map of the source
intensity distribution by this number.

The ALCs are employed to keep the input power to the correlator constant, which
in turn keeps the effective gain of the correlator constant. However, our tests
show that a variation in input power by a factor of 4 changes the effective
gain of the 4-bit correlator by $\le 5$\%. Since the variation in total output
power between the GC and the secondary calibrator with the ALC off is also
$\sim 4$, to prevent $T_{sys}$ affecting directly the visibility amplitudes,
the ALCs were turned off for 580 MHz and 1010 MHz observations, which were
carried out after the 620 MHz observations. The absolute flux density scale
using the first method (i.e., at 620 MHz) is believed to be accurate to about
10\%, while, it is expected to be accurate to better than 5\% with ALCs off
(i.e., at 580 and 1010 MHz).

The data were processed within the Astronomical Image Processing System
(AIPS) using standard programs. Bad data were identified and flagged using
various tasks in AIPS.
After calibration and editing of the 620 MHz data, a pseudo-continuum database
of 3 frequency channels (each of width 3.7 MHz) was made from the central 11
MHz of the observed 16 MHz band. This was adequate to avoid bandwidth smearing
within the primary beam. Images of the fields were formed by Fourier inversion
and Cleaning (IMAGR).  The initial images were improved by phase only
self-calibration (Self-cal).  However, subsequent amplitude and phase Self-cals
failed to improve the dynamic range of the maps. Therefore, phase only
Self-cals were used to produce the final images.

\begin{figure}
\centering
\includegraphics[width=0.45\textwidth, clip=true,]{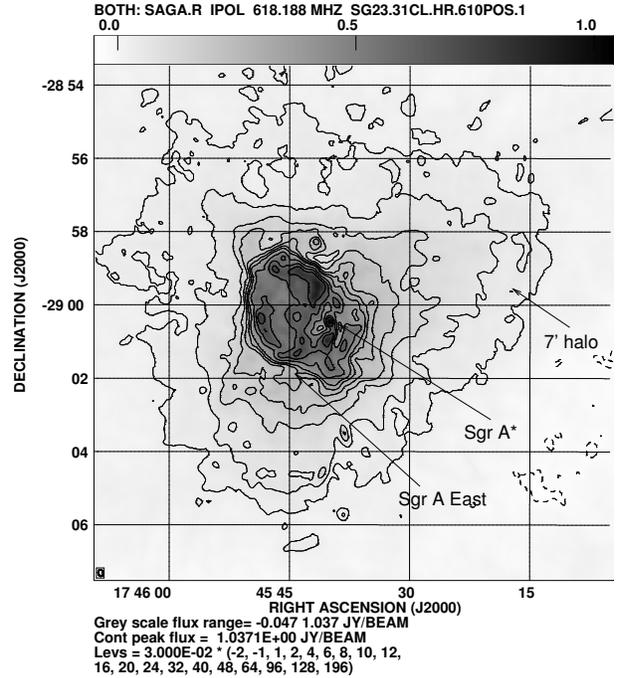} \caption{620 MHz GMRT map of the central 15$'$ region of
the Galaxy (in both contour and gray scale). The resolution is 11.4$^{''}$
$\times$ 7.6$^{''}$, with beam position angle of 7.7\dg. The rms noise is about
6.5 mJy/Beam.}
\label{620mhz.gc.central.map}
\end{figure}

To improve the deconvolution of the extended emission, we made the final
image (Fig.~\ref{620mhz.gc.central.map}) using Multi-resolution Clean
\citep{WAKKER1988} as implemented in AIPS.  Since the strong emission is
seen mostly near the central 10$'$ region of the GC, and we are interested
in sources located within only this region, no 3D deconvolution of the
dirty image to compensate for the non-coplanar baseline effect was
performed.
The GMRT map at 620 MHz (Fig.~\ref{620mhz.gc.central.map}) has a dynamic
range of about 150, and is limited by systematics. The above procedures
were also employed to calibrate the 580 MHz and 1010 MHz data, and to make
images at 580 MHz.  Phase only Self-cals were performed with the Clean
Components of Sgr~A* at 1010 MHz to produce the final self-calibrated {\it
uv}-data.

\begin{figure}
\centering
\includegraphics[width=0.46\textwidth, angle=0,clip=true]{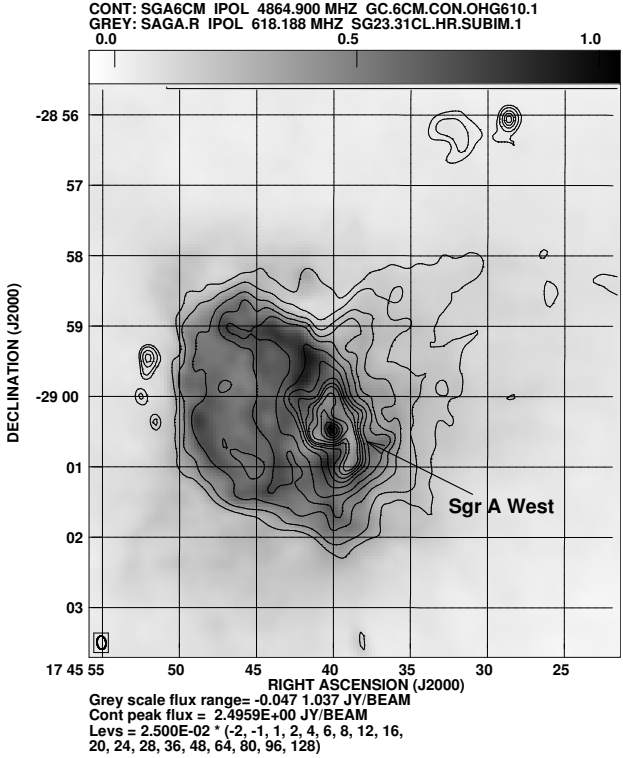}
\caption{4.8 GHz continuum map of the Sgr A complex \citep{YUSEF-ZADEH1989} in
contour overlaid on the 620 MHz gray scale map of the same region. The
resolution is 11.4$^{''}$ $\times$ 7.6$^{''}$, with beam position angle of
7.7\dg. The rms noise in the 4.8 GHz map is about 4 mJy/Beam.}
\label{4.8ghz.49cm.gc.central.map}
\end{figure}

\section{Results}

\subsection{Features in the 620 MHz map}
The central 15$'$ region of the Galaxy at 620 MHz is shown in
Fig.~\ref{620mhz.gc.central.map}. The compact source Sgr~A* is clearly seen,
along with other well known sources like Sgr A East and the 7$'$ halo, which
were also seen in our earlier 610 MHz map \citep{IAU199.ROY}, but due to change
in calibration scheme and in subtraction of the background extended emission
(see Sect.~\ref{sgra*.flux}), the earlier results from \citet{IAU199.ROY} will
not be used in drawing any quantitative estimation in this paper.

To compare the smaller scale features near Sgr~A~West with what is seen at
higher frequencies, we have plotted in Fig.~\ref{4.8ghz.49cm.gc.central.map},
the 620 MHz map in gray scale and the 4.8 GHz VLA map in contours convolved to
a common resolution. There is almost one to one correspondence between the
higher emission features at 4.8 GHz comprising the Sgr~A~West region and a drop
in the total intensity at 620 MHz (indicated by white region in the gray scale
map), indicating that the thermal emission is optically thick near 620 MHz. The
shell-like emission feature known as Sgr A East is clearly identified in both
the maps. An emission feature $\approx$30$^{''}$ south of Sgr~A* can be seen in
the 620 MHz gray scale map.  This feature was identified by \citet{PEDLAR1989},
who suggested that it is associated with Sgr~A~East. The Sgr A East shell has a
size of about 2$' \times$ 3$'$, and there is a triangular shaped halo of size
$\sim 7'$ around its shell, which is known as the 7$'$ halo.

\begin{figure}
\centering
\includegraphics[width=8.0cm,angle=0,clip=]{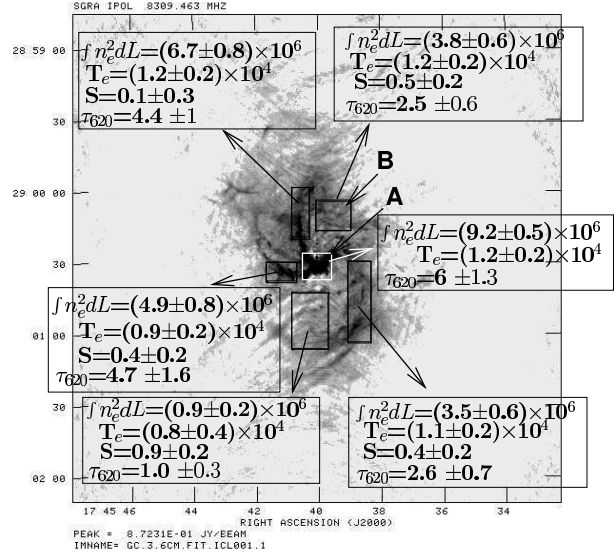}
\caption{Estimated electron temperature, emission measure (in boxes) and
average optical depth at 620 MHz ($\tau_{620}$) at 6 different parts of
Sgr~A~West. These parts are enclosed in rectangular boxes drawn on the 8.3 GHz
gray scale map \citep{ROBERTS1993}}
\label{sgra.west.fit}
\end{figure}

\subsection{Emission measure and temperature of the ionised gas in the
Sgr~A~West HII region}

In this section, we estimate the density and temperature of ionised hydrogen in
Sgr~A~West. \citet{PEDLAR1989} estimated these parameters by assuming a
constant electron temperature across Sgr~A~West. However, due to the
availability of high resolution data at 4 different frequencies between 8.3 GHz
and 620 GHz, we can study the variation of the electron temperature (T$_e$),
the emission measure ($\int n_e^2 dL$, where $L$ is the path length through the
ionised medium) and the background non-thermal flux density due to Sgr A East
as functions of position in the map. To estimate the quantities, the optical
depth of the Sgr~A~West HII region as a function of frequency needs to be
estimated, which we describe below.

While estimating flux densities from different maps, the synthesised beam of
each of the maps needs to be equal. We have convolved the 8.3 GHz, 4.8 GHz, 1.4
GHz and 620 MHz maps to a common resolution of 11.4$^{''}$ $\times$ 7.6$^{''}$,
with beam position angle of 7.7\dg. In Fig.~\ref{620mhz.gc.central.map}, there
are four sources, Sgr~A*, Sgr~A~West, Sgr A East and the 7$'$ halo, which
appear to overlap along the line of sight. Because of this overlap, estimating
flux densities of Sgr~A~West is non-trivial.
Based on the suggestion of \citet{PEDLAR1989}, we assume that the 7$'$ halo is
located in front of the Sgr A complex. We took crosscuts at several
orientations across the 7$'$ halo through the location of Sgr~A*, and estimated
its mean contribution across the Sgr~A complex.
The estimated intensity of the 7$'$ halo is 0.18 Jy/Beam at 620 MHz and at 1.4
GHz.
The background emission due to 7$'$ halo is quite small at 4.8 GHz, where it
gets partially resolved due to the lack of short interferometric spacings, and
we estimate its contribution at this frequency to be about 0.012 Jy/Beam (no
background subtraction was performed at 8.3 GHz). After subtracting these
intensities of the 7$'$ halo from the corresponding maps, the flux densities at
different parts of Sgr~A~West were estimated. Even if the 7$'$ halo is
considered to be a mixture of thermal and non-thermal emission
\citep{PEDLAR1989}, from the estimates made by \citet{PEDLAR1989} at 330 MHz,
it can be considered to be optically thin at frequencies of 620 MHz and above.
Consequently, after subtraction of emission from the 7$'$ halo and the removal
of the contribution from Sgr~A*, the observed flux density towards Sgr~A~West
is the sum of non-thermal emission from Sgr A East (located behind Sgr~A~West
(Pedlar et al. 1989)) and its own thermal emission.
The observed flux density from Sgr A East can be expressed as $ S \nu^{\alpha}
\times exp(-\tau)$, where $S$ is the flux density of the background source at 1
GHz in Jy/Beam, $\tau$ is the free-free optical depth at a frequency of $\nu$
GHz and $\alpha$ is the spectral index. The estimated flux density due to
free-free emission is [2 $k T_e \times (1 - \exp(-\tau))\times \nu^2 \times
\Delta\Omega]/c^2$.  Where, $\Delta\Omega$ is the synthesised beam size. From
the radiative transfer equation, we can express the estimated flux density per
synthesised beam as 
\begin{equation}
I_{\nu}=S \nu^{\alpha} \times \exp(-\tau) + [2 k T_e \times (1 - \exp(-\tau))
\times \nu^2 \times \Delta\Omega]/c^2.
\end{equation}

Where, $ \tau = \int 0.2 n_e^2 T_e^{-1.35} \nu^{-2.1} dL $.

Following \citet{PEDLAR1989}, we assume the spectral index of Sgr A East to be
$-$1.0 (S($\nu$) $\propto \nu^{\alpha}$) between 8.3 GHz and 620 MHz. The
average intensity per synthesised beam estimated at different frequencies from
each of 4 different regions corresponding to the Western arc, Northern and
Southern arm, and the diffuse halo (sampled at two different parts) as shown
in Fig.~\ref{sgra.west.fit}, were least square fitted with the expression for
$I_\nu$ given above. The estimated T$_e$ and $\int n_e^2 dL$ and the optical
depth at 620 MHz ($\tau_{620}$) for these regions are also shown in
Fig.~\ref{sgra.west.fit}. We note that the estimated parameters do not vary
significantly, even if the spectral index of Sgr A East changes from $-$1.0 by
$\pm$0.5 between 1.0 GHz and 620 MHz. Sgr~A* is seen through the region marked
`A', and we subtracted its observed flux density at the position of Sgr~A* (see
below) from each of the different frequency maps. The fit of intensities of
region `A' to the expression for $I_\nu$ did not converge, and `$S$' was held
fixed to zero to get a fit (see Fig.~\ref{sgra.west.fit}). Since there is no
measurable increase in the flux density from this region between 8.3 and 4.8
GHz (thermal emission is optically thin at these frequencies), the non-thermal
emission from Sgr A East is negligible, and therefore keeping $S$ fixed to
`zero' should have negligible effect on the estimated parameters.

From the model fit to all of the above mentioned regions, the electron
temperature is found to be within 8,000--12,000 K, which is higher than the
reported value of 7000~K by recombination line observation of
\citet{ROBERTS1993}. They assumed the emitting gas to be in local thermodynamic
equilibrium (LTE), and that pressure broadening is negligible. However, the
estimate of T$_e$ from the continuum spectrum, anchored at the low frequencies
by the GMRT measurements, is free from any such assumptions and provide a more
accurate value.

\subsubsection{Flux density of Sgr~A*}
\label{sgra*.flux}
While estimating the flux density of Sgr~A* at 620 MHz, we reduced the
confusion due to the extended emission by imaging only those visibilities
having {\it uv} distance $>$7 $k\lambda$. The flux density estimated from the
image plane is about 0.5 $\pm$0.1 Jy. We note that even after applying a lower
{\it uv} cutoff, there is significant background confusion within a beam of
size 7.5$^{''} \times 4^{''}$. This confusion causes an uncertainty of about
0.1 Jy in the estimated flux density. We also estimated the flux density of
this object from the {\it uv} visibilities. We first applied appropriate phase
shift such that Sgr~A* is at the phase centre. Every one hour of {\it uv}-data
were averaged vectorially, which ensures that the sources away from the centre
have little contribution to the visibility amplitude.  After rejecting data
with {\it uv}-distance shorter than 15 k$\lambda$, an elliptical Gaussian model
was fitted as a model for Sgr~A* (using UVFIT).  The flux density estimated in
this way is also 0.5 Jy.  The major and minor axis of the Gaussian fit is
3.8$''\pm$0.4$''$ and 1.8$''\pm$0.6$''$ respectively with a position angle of
93$\pm$4\dg, which is consistent with its expected scatter broadened size of
3.4$'' \times 1.8''$ estimated from \citet{LO1998} at this frequency. To check
the goodness of the fit, we divided the {\it uv}-data by the model. Similar
broadening of compact maser sources in the vicinity of OH/IR stars have been
observed in the central 30$'$ of the Galaxy \citep{VANLANGEVELDE1992}. The real
part of the data after division indeed is close to unity with a scatter of less
than 0.2, which indicates an rms error of 0.1 Jy in the flux density
measurement.

We also imaged Sgr~A* from the 580 MHz data. The flux density estimated at this
frequency matches (within the error-bar) to what is estimated at 620 MHz and
its known spectrum at radio frequencies. The flux density of Sgr~A* at 1010 MHz
as estimated from the {\it uv} visibilities is 0.6 $\pm$0.12 Jy.

\begin{figure}
\centering
\includegraphics[width=0.36\textwidth, angle=270,clip=true]{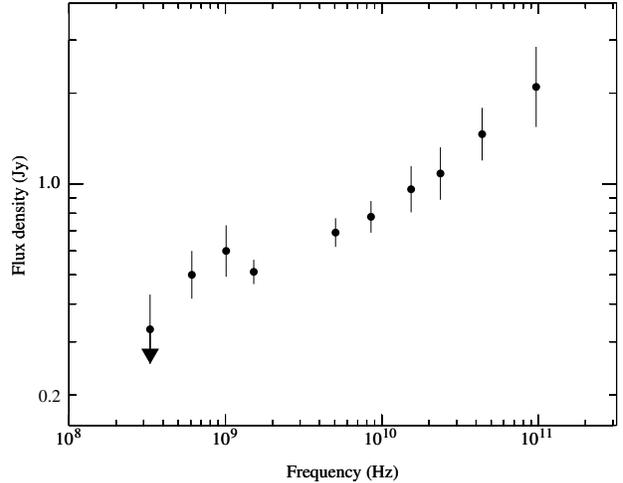}
\caption{The observed spectrum of Sgr~A* from 300 MHz to 20 GHz. Except the
330, 620 and 1010 MHz measurements, all the other data points are taken from
\citet{MELIA2001}. In the plot, the 330 MHz flux density estimated at the
position of Sgr~A* \citep{NORD2004} is taken as the upper limit.}
\label{sgra*.spectrum}
\end{figure}

\section{Discussion}
\subsection{Low frequency spectral index of the Sgr~A*}

Although the high radio frequency spectrum of Sgr~A* is well established, the
spectrum below 1.4 GHz is not well estimated. At 1010 MHz, the flux density of
Sgr~A* is about 0.6 Jy and its time averaged spectral index between 1.4 and 8.5
GHz is +0.17 \citep{MELIA2001}. \citet{DAVIES1976} found the flux density of
Sgr~A* at 960 MHz to be a factor of 2 less than at 1.6 GHz and suggested that
it has a low frequency turnover around 1 GHz. This appears to be confirmed from
their upper limit of 50 mJy at 408 MHz and the 100 mJy upper limit set by
\citet{PEDLAR1989} at 330 MHz. However, \citet{NORD2004} at 330 MHz estimate a
total flux density of 330 $\pm 120$ mJy from the known position and expected
size of Sgr~A*.  The recent flux density estimates of Sgr~A* at 1010, 620 and
330 MHz raises questions about earlier measurements. Based on the average 
flux density of Sgr~A* at 1.4 GHz \citep{ZHAO2001} and its known spectrum
between 1.4 and 8.5 GHz, we expect a flux density of 0.47 Jy at 1010 MHz and
0.44 Jy at 620 MHz, which are consistent with our measurements within the
error-bars.
The observed spectrum of Sgr~A* from 300 MHz to 20 GHz is shown in
Fig.~\ref{sgra*.spectrum}. From the spectrum it is apparent that a factor of
two lower flux density at 960 MHz \citep{DAVIES1976} can be explained only by
assuming Sgr~A* to be variable at this frequency. \citet{FALCKE1999} finds a
modulation index of 6\% at 8.3 and 2.5\% at 2.3 GHz, which indicates that the
intrinsic variability of Sgr~A* is rather low at low radio frequencies
\citep{MELIA2001}. Therefore, interstellar scintillation (ISS), the time scale
of which is $\sim$few years towards Sgr~A* at 1 GHz, is likely to cause any
significant variability.  The observations in the 620 MHz band cover a span of
nearly an year and no significant variation ($\sim 20$\%) could be detected.
Our measurements rule out any turnover near 1 GHz and indicate that any
turnover has to be at frequencies less than 580 MHz. The upper limits at 408
MHz, however, pose problems for this picture, as this would suggest the flux
density of Sgr~A* at this frequency could have been much less few decades back.
We have re-examined the 408 MHz upper limit by \citet{DAVIES1976} and found
that their analysis has not taken account of scatter broadening of Sgr~A*
(about 8$''$ at 408 MHz) and the corrected upper limit at 408 MHz could be as
high as $\approx$~2.5 Jy.

At frequencies of a few GHz to 620 MHz, the observed spectral index of Sgr~A*
is nearly flat. For extragalactic flat spectrum sources, it is believed that
there are multiple components in emission and peak radiation from different
components occur at different frequencies to make the overall spectrum appear
nearly flat \citep{COTTON1980}.  Though the Advection-dominated accretion flow
(ADAF) model of emission from Sgr~A* fails to explain the observed low
frequency emission, ADAF along with self-absorbed synchrotron emission either
from a relativistic Jet \citep{YUAN2002}, or from a small fraction of
relativistic electrons embedded in the accretion flow \citep{YUAN2003} could
explain the observed spectrum of Sgr~A* from X-ray ranges to radio frequencies
up to 620 MHz. If we consider the flux density estimated at the position of
Sgr~A* at 330 MHz \citep{NORD2004} as the observed upper limit, its estimated
spectral index between 620 and 330 MHz is inverted and is $>$0.66.  This low
frequency turnover could be due to an enhancement in the optical depth of
synchrotron emission or internal/external free-free absorption.

\subsection{Location of Sgr~A*}

At 620 MHz, Sgr~A~West shows evidence for free-free self-absorption. In
Fig.~\ref{sgra.west.fit}, we have shown the mean optical depth for 6 different
regions of Sgr~A West. The Sgr~A* being slightly above the junction of the Bar
and the Northern arm \citep{ROBERTS1993}, optical depth near Sgr~A* is almost
the same as the Sgr~A West halo marked `B', and is 2.5$\pm$0.5.
If Sgr~A* was located behind Sgr~A~West, then its flux density would have been
attenuated by a factor of 10.  However, the spectral index of Sgr~A* between
620 MHz and 1.4 GHz is roughly the same as between 1.4 GHz and 8.5 GHz showing
no effect of free-free absorption by Sgr~A~West. This indicates that Sgr~A* is
located in front of Sgr~A~West. It is possible to invoke alternate scenarios
like sharp enhancement of the 620 MHz flux density of Sgr~A* to compensate for
the absorption due to Sgr~A~West or a hole in Sgr~A~West along the line of
sight to Sgr~A*, but these appear to be unlikely.  In a high resolution radio
continuum (resolution $\sim 0.6^{''}$) map \citep{ROBERTS1993}, Sgr~A* appears
to be located slightly ($\sim 1^{''}$) above the junction of the Bar and the
Northern arm and diffuse emission can be seen near its position. In radio
recombination line (resolution $\sim 2^{''}$) data \citep{ROBERTS1993},
emission at the position of Sgr~A* is observed near a velocity of 56 \kms.
There is a region of molecular line emission at the outer edge of the Sgr~A
West, which is known as Circumnuclear Disk (CND), and \citet{ROBERTS1993}
suggests the recombination line to be stimulated emission from a region near
the interior edge of the CND. However, the emission is observed beyond the
synthesised half power beam width of Sgr~A* and the velocity of emission
matches with that of the H~92$\alpha$ emission from Northern arm. Therefore, it
appears that the diffuse gas seen in the radio continuum image near Sgr~A* is
actually associated with the Northern arm (see also \citet{SCHWARZ1989}), which
indicates that it is located behind Sgr~A* (see also Sect.~1).  Any hole in
this ionised gas has to be smaller than $1''$, which is unlikely.  Thus, Sgr~A*
is located in front of Sgr~A~West.

\section{Conclusions}
Observations of the GC region at 1010, 620 and 580 MHz with the GMRT and a
comparison with the existing observations made at other radio frequencies have
provided us several important details about the region:\\
(i) Sgr~A* has been detected at 580 MHz, which is the lowest frequency
unambiguous detection of Sgr~A*, and the estimated flux density is consistent
with what is expected from its higher radio frequency spectral index and the
flux density.  This indicates that there is no low frequency turnover of its
emission at freqencies above 0.6 GHz. \\
(ii) Our observations at 0.6 GHz breaks the degeneracy between emission measure
and electron temperature in the existing data, and allows us to estimate the
optical depth of the Sgr~A~West HII region to be about 2.5 at 620 MHz. \\
(iii) Though Sgr~A* is located along the same line of sight as Sgr A West, the
emission from it undergoes no absorption by this HII region, which indicates
that Sgr~A* is located in front of Sgr~A~West.

\section*{Acknowledgements}
S.R. thanks Miller Goss and Jun Hui Zhao for useful discussions. We also thank
Rajaram Nityananda for reading the manuscript and for making useful comments.
We thank the staff of the GMRT that made these observations possible. GMRT is
run by the National Centre for Radio Astrophysics of the Tata Institute of
Fundamental Research. We have used FITS images from the NCSA Astronomy
Digital Image Library (ADIL), and the authors wish to thank them.

\bibliographystyle{mn2e}
\bibliography{gc.pap}
\label{lastpage}
\end{document}